\documentclass[preprint,aps]{revtex4}
\begin{document}
\title{ Two-capacitor problem revisited: A mechanical harmonic oscillator model approach
\\}
\author{Keeyung Lee \\
}
\address{
Department of Physics, Inha University,
Incheon, 402-751, Korea\\
}
\date{\today}
\begin{abstract}
The well-known two-capacitor problem, in which exactly half the
stored energy disappears when a charged capacitor is connected to
an identical capacitor is discussed based on the mechanical harmonic
oscillator model approach. In the mechanical harmonic oscillator model,
it is shown first that \emph {exactly half} the work done by a constant applied force is dissipated
irrespective of the form of dissipation mechanism when the system comes to
a new equilibrium after a constant force is abruptly applied. This model is
then applied to the energy loss mechanism in the capacitor charging problem or the two-capacitor problem.
This approach allows a simple explanation of the
energy dissipation mechanism in these problems and shows that the dissipated energy
should always be \emph {exactly half} the supplied energy
whether that is caused by the Joule heat or by the radiation.

\end{abstract}
\narrowtext
\maketitle
\vfil\eject

{\bf 1. INTRODUCTION}

When a charged capacitor is connected to another identical one,
exactly half the stored energy disappears after the charge
transfer is completed. Such situation also occurs when a capacitor
is charged by a battery, where only half the supplied energy is
left as the stored energy in the capacitor after the charging
process is completed. This problem has a long history and it is an
interesting problem pedagogically\cite{cuvaj,powell,boykin,choy}.

The question of 'what is the mechanism of such energy loss, and why \emph {exactly half} the supplied energy is
lost' has been the point interest in this problem. It was noted in earlier works, that if the presence of nonzero
resistance connecting the two capacitor is assumed, the amount of Joule heat loss in the resistor becomes exactly
the same as the 'missing' energy\cite{cuvaj}. Such explanation shows that resistance is an indispensable part in
any electrical circuit. (One can imagine using the superconducting wire for the connection, but it is obvious
that the increasing current will quickly reach the critical limit which will destroy the superconducting state.
Therefore, this example shows that critical field limit must exist in the superconducting phenomena.)

Although this problem was resolved by such a simple model, of course such an $RC$ circuit model is not realistic.
Therefore Powell has considered a more realistic model where an inductor element is included, since any closed
circuit has to have self-inductance effect\cite{powell}. It was shown that in all three cases of overdamped,
underdamped and critically damped cases, Joule heat loss by the resistor becomes the same, accounting for the lost
energy exactly. Powell noted that there are radiation effects which may be important. Therefore although the
missing energy problem was resolved, it was not clearly understood why the simple $RC$ model was as satisfactory
as the more realistic $RLC$ model, and why discussions which do not incorporate the radiation effect could also
give perfect explanations.

We discuss this problem based on the simple harmonic oscillator
model approach in this work. From this approach it is shown that
the missing energy should always be exactly half whether the
dissipation mechanism is the Joule heat loss or the radiation
effect of the combination of them. For this purpose, motion of an
object in a harmonic potential under the influence of damping
force which is velocity and acceleration dependent is considered.
Our analysis does not require the exact solution of differential
equation which may be difficult or impossible to obtain for such
cases.

\bigskip
 {\bf 2. CLASSICAL HARMONIC OSCILLATOR MODEL}

Let us begin our discussion by considering a simple harmonic
oscillator where a frictional force of the type $f(v, dv/dt)$
which depends on the velocity $v$ and acceleration $dv/dt$ is
present. If a constant external force $F_0$ is applied to an
object of mass $m$ in a harmonic potential of the form $(1/2)
kx^2$, the equation of motion can be written as follows,

$$ m{dv \over dt} = F_0  - f(v, dv/dt) - kx \eqno(1)$$

Analytic solution of this equation cannot be obtained in general
except for some simple special cases. Let us assume that the
constant force $F_0$ is applied at time $t=0$, and the object is
at rest at the equilibrium position at that instant. The object
will come to a complete stop at the new equilibrium position
eventually, therefore the initial and final velocity are both 0,
which means the resulting change of the kinetic energy becomes 0.

Let us first consider the energy dissipation aspect of this system when the damping force term does not depend on
acceleration and is simply proportional to the velocity such that $f(v, dv/dt) = bv$, in which $b$ is a constant.

Let the final equilibrium position of the object be denoted as
$x_0$. Since both the velocity and acceleration of the object
become 0 eventually, as long as the damping force disappears when
$v=0$ and $dv/dt=0$, which is naturally satisfied when the object
comes to a complete stop, it could be noted that the final object
position can still be expressed as $x_0 = F_0 /k $. Therefore the
resulting change of potential energy can be written as $ (1/2)
kx_0 ^2$. If these facts are used, the integrated result of this
equation can be expressed as,

$$ F_0{\int_0 ^\infty v dt } = {\int_0 ^\infty  f(v,dv/dt) dt} + {{1 \over 2} kx_0 ^2} \eqno(2)$$

The left hand side of the equation indicates the amount of work
done by the force and the first term on the right hand side
indicates the amount of dissipated energy. Although $v(t)$ may
have a complicated form, it is important to note at this point
that $ \int_0 ^\infty v dt $ can be identified as the eventual
position $x_0$ of the object. Since $x_0 = F_0 /k $, this means
that the amount of supplied work can be expressed as $kx_0 ^2 $.
Therefore the following interesting relation can be obtained.

$$ {kx_0 ^2} = {\int_0 ^\infty bv^2 dt} + {{1 \over 2 } kx_0 ^2} \eqno(3) $$

This relation shows that {\emph {exactly half} the work done by
the applied force is eventually lost as the dissipated energy. Let
us call this relation as, "dissipation energy relation in a
harmonic potential system". This relation can also be proved using
the analytic solutions (which are usually classified as
overdamped, underdamped, and critical damped cases) for this
particular case.

It should be noted again that this relation has been obtained
without resorting to the solution of differential equation.
therefore we are allowed to have a general form of the damping
force. Let us assume that the damping force function has a
polynomial form of the type $f(v) = b_1 v + b_2 v^2 +...$. It is
known that, in the case of viscous resistance on a body moving
through a fluid, such a form has to be assumed\cite{fowles}. The
equation of motion in this case becomes nonlinear and it is
usually impossible to obtain the exact solution of such type of
equations, but of course, this dissipation energy relation still
holds. An interesting point is that, this is true even when the
damping force function depends on the acceleration. This fact will
be used as an essential point in the following two-capacitor
problem discussion.

In some physical systems, potential function may include
anharmonic terms such as $kx^n$, where $n$ is an integer not equal
to 2, which could lead to nonlinear equation of motion.
(Anharmonic potential term is important in describing the
intermolecular potential in molecular bonding.) The exact solution
of differential equations for these cases is usually not
available. But it can be noted that a definite relation between
the supplied and dissipated energy can still be found based on our
analysis, although the dissipation energy relation for these cases
will become different from the simple harmonic systems.

\bigskip
{\bf 3. DISCUSSION OF THE TWO-CAPACITOR PROBLEM }

Let us now extend our discussion to the two-capacitor problem. For this purpose, we consider an $RLC$ circuit
with resistance $R$, and inductance $L$ connected to an {\it emf} source of magnitude $\epsilon_0$ as in Fig. 1(a), which has the following circuit equation,

$$ {\epsilon_0} - R i - L {di \over dt} - {1 \over C} q  = 0 \eqno(4)$$

If the current $i$ and inductance $L$ are considered to correspond
to velocity $v$ and mass $m$ in the mechanical system, this
equation is the exact equivalent of the damped harmonic
oscillator. It could also be shown that this equation is
essentially equivalent to the two-capacitor problem
equation\cite{powell}.

Now if the current $i$ is multiplied to this equation, we obtain,

$$ {\epsilon_0} i- Ri^2 - {dU_m \over dt} - {dU_e \over dt} = 0 \eqno(5)$$

where $U_m = {1 \over 2} L i^2$, and $U_e = {1 \over 2} {1 \over C}q^2$. If this equation is integrated,
then the following relation can be obtained.

$$ {\epsilon_0} {\int_0 ^\infty} i dt = {\int_0 ^\infty } Ri^2 dt + {\Delta U_m} |_0 ^{\infty} + {\Delta U_e} |_0 ^{\infty}  \eqno(6)$$

Since the current $i$ is equal to 0 at $t=0$ and at $t=\infty$, $ \Delta U_m $ part vanishes. Also since both $i$
and $di/dt$ vanish eventually, the amount of capacitor charge become $ C \epsilon_0$ from Eq.(4), and therefore
$\Delta U_e$ part becomes  $(1/2) C \epsilon_0 ^2$ and the following expression can be obtained.

$$ C {\epsilon_0 ^2} = {\int_0 ^\infty} Ri^2 dt + {1 \over 2} C {\epsilon_0 ^2} \eqno(7)  $$

This shows that only half the supplied energy is left as the
stored energy in the capacitor and the other half is necessarily dissipated in
the resistor whenever one attempts to charge a capacitor.

Now consider the circuit in which an initially charged capacitor with charge $Q_0$ is connected to another capacitor of the same capacitance as in Fig. 1(b). If the charge of the connected capacitor is denoted as $q$, the circuit equation for this case can be written as,

$$ {{Q_0 -q}\over C} - R i - L {di \over dt} - {1 \over C} q  = 0 \eqno(8)$$

If we replace $Q_0 / C $ as $ \epsilon_0$, this can be rewritten as,

$$ {\epsilon_0} - R i - L {di \over dt} - {2 q\over C}   = 0 \eqno(9)$$

which is the same as the Eq.(4), except that the capacitor has been replaced by one with half the capacitance of $C/2$.

Following the analysis as has been done, considering now that ${\int_0 ^\infty} i dt = Q_0 /2 $ and that eventually the capacitor is charged to $Q_0 /2$, one obtains the relation,

$$ {1\over 2 } C {\epsilon_0 ^2} = {\int_0 ^\infty} Ri^2 dt + {1 \over 4} C {\epsilon_0 ^2} \eqno(10)  $$

This shows that the total dissipated energy become equal to $ {1\over 4} Q_0 ^2 / C $, which is just half the originally stored energy. In fact, the circuit equations for the capacitor charging case by a battery as is given by Eq.(4) and the two-capacitor problem case as is given by Eq.(8) become the essentially the same if differentiated, which means that two situations are basically identical.

Therefore we could successfully resolve the
two-capacitor problem in a much simpler way than what Powell did,
i.e. without resorting to the solution of differential equation.
In fact, what Powell did was to find solutions for the
differential equation, and prove this result by evaluating the
integrals directly\cite{powell}. Powell has found that the
overdamped case closely approximates the less realistic $RC$
situation, which may explain why the unrealistic $RC$ model could
also be successful. But the interesting point is that $RC$ model
explains the missing energy exactly, not approximately.
Furthermore, it was found that most practical circuits belong to
the underdamped case, not the overdamped case. Therefore the
reason for the success of $RC$ model could not be understood. In
the present analysis, since $\Delta U_m = {1 \over 2} L i^2$ part
vanishes anyway, it is easy to see why the presence of inductor in
the circuit does not make any difference at all for this problem.

Although the circuit including the inductor component is more
realistic, it is not truly realistic because it does not consider
the radiation effect which becomes especially important in high
frequency oscillating circuits. To resolve this situation, Boykin
{\it et al.} have adopted the magnetic dipole model and have shown
that just the radiation effect due to the magnetic dipole model
can also explain the missing energy\cite{boykin}. Using the so
called "lumped-parameter model" (in which nonlinear circuit
elements excluding the capacitor radiation effect have been
represented by an equivalent radiation resistance), it was shown
that the presence of resistor is not essential to resolve the
two-capacitor problem.

There exists oscillating electric field inside the capacitor which can also radiate. Such electric dipole
radiation effect from the capacitor part was also considered by Choy, who showed that the electric dipole model
can also explain the missing energy\cite{choy}. It is amazing to find that all approaches from the simple $RC$
model to the more elaborate radiating dipole models could all successfully resolve the two-capacitor problem. But
the reason for such outcome could not be understood as yet.

To understand this situation, let us now extend the above model
which corresponds to the velocity dependent damping force model to
a case which includes acceleration dependent damping force. It is
well known that the radiation effect can be incorporated as a
resistive term\cite{jackson}. Since the dipole radiation effects
involve the acceleration term, our resistive force function is now
considered to have a form of $f(v, dv/dt)$. However, we have
already found that our dissipation energy relation remains
effective even for such form of resistance functions. Therefore it
could be stated that regardless of the nature of damping force,
i.e., whether that is the Ohmic resistive force or radiative
resistive force or the combination of both, the amount of
dissipation energy should always be exactly half the supplied
energy in the two-capacitor problem.

\bigskip
{\bf 4. CONCLUSION}

The energy loss mechanism in the two-capacitor problem has been
considered using a harmonic oscillator model in this work. We have
shown, without resorting to solutions of differential equations,
that the amount of energy loss in the form of Joule heat should be
exactly half the supplied energy in the $RLC$ circuit. Since our
analysis does not require the solution of differential equation,
we could extend our analysis to damping forces which include an
acceleration term, which corresponds to radiative resistance. This
explains why perfect explanations of the two-capacitor problem
have been possible whether just the Joule heat loss or dipole
radiation energy loss is considered, providing a comprehensive
understanding of the two-capacitor problem.


\begin{references}
%
\bibitem{cuvaj} C. Cuvaj, "On conservation of energy in electric circuits", Am. J. Phys., {\bf36}, 909-910 (1968)
\bibitem{powell} R.A. Powell, "Two capacitor problem: A more realistic view", Am. J. Phys., {\bf47}(5), 460-462 (1979)
\bibitem{boykin} T.B. Boykin, D. Hite, and N. Singh, "The two-capacitor problem with radiation", Am. J. Phys., {\bf70}(4), 415-420 (2002)
\bibitem{choy} T.C. Choy, "Capacitors can radiate: further results for the two-capacitor problem", Am. J. Phys., {\bf72}(5), 662-670 (2004)
\bibitem{fowles}  Fowles, {\it Analytical Mechanics} (Wiley, New York, 1975) 2nd ed. p 783.
\bibitem{jackson} J.D. Jackson, {\it Classical Electrodynamics} (Wiley, New York, 1975) 2nd ed. p 783.
\end{references}
\end{document}